\documentclass[aip,graphicx, rsi, amsmath,amssymb,preprint]{revtex4-1}
\usepackage{graphicx}
\usepackage{dcolumn}
\usepackage{bm}
\usepackage{xcolor}
\usepackage{subcaption}

\usepackage[utf8]{inputenc}
\usepackage[T1]{fontenc}
\usepackage{mathptmx}

\draft 

\begin{document}


\title{\textit{Ab initio} metadynamics determination of temperature-dependent free-energy landscape in ultrasmall silver clusters} 


\author{Daniel Sucerquia}
\affiliation{Biophysics of Tropical Diseases, Max Planck Tandem Group, University of Antioquia UdeA, 050010 Medellin, Colombia
}%
\affiliation{Grupo de Física Atómica y Molecular, Instituto de Física, Facultad de Ciencias Exactas y Naturales, Universidad de Antioquia UdeA; Calle 70 No. 52-21, Medellín, Colombia}

\author{Cristian Parra}%
\affiliation{Biophysics of Tropical Diseases, Max Planck Tandem Group, University of Antioquia UdeA, 050010 Medellin, Colombia
}%

\author{Pilar Cossio}
\email{pcossio@flatironinstitute.org}
\affiliation{Biophysics of Tropical Diseases, Max Planck Tandem Group, University of Antioquia UdeA, 050010 Medellin, Colombia
}%
\affiliation{Center for Computational Mathematics, Flatiron Institute, NY, USA.
}%

\author{Olga Lopez-Acevedo}
\email{olga.lopeza@udea.edu.co}
\affiliation{Biophysics of Tropical Diseases, Max Planck Tandem Group, University of Antioquia UdeA, 050010 Medellin, Colombia
}%
\affiliation{Grupo de Física Atómica y Molecular, Instituto de Física, Facultad de Ciencias Exactas y Naturales, Universidad de Antioquia UdeA; Calle 70 No. 52-21, Medellín, Colombia}

\date{\today}

\begin{abstract}
\textit{Ab initio} metadynamics enables extracting free-energy landscapes having the accuracy of first principles electronic structure methods. We introduce an interface between the PLUMED code that computes free-energy landscapes and
enhanced-sampling algorithms and the ASE module, which includes several \textit{ab initio} electronic structure codes. The interface is validated with a Lennard-Jones cluster free-energy landscape calculation by averaging multiple short metadynamics trajectories. We use this interface and analysis to estimate the free-energy landscape of Ag$_5$ and Ag$_6$ clusters at 10, 100 and 300 K with the radius of gyration and coordination number as collective variables, finding at most tens of meV in error. Relative free-energy differences between the planar and non-planar isomers of both clusters decrease with temperature, in agreement with previously proposed stabilization of non-planar isomers.  Interestingly, we find that Ag$_6$ is the smallest silver cluster where entropic effects at room temperature boost the non planar isomer probability to a competing state. The new ASE-PLUMED interface enables simulating nanosystem electronic properties at more realistic temperature-dependent conditions.  
\end{abstract}


\maketitle 

\section{Introduction}

 Metadynamics (MTD) is a free-energy estimation method that enables exploring the conformational space of a system at a given temperature. It relies on the theoretical relation between the free energy of the system and a bias potential that drives the system to cross barriers and explore new conformations \cite{Bussi2006, Barducci2008}. In principle, such algorithms can be coupled to any energy-force level description of the system. However, most applications prefer classical to quantum methods. Metadynamics with quantum methods has been used to simulate chemical and biochemical reactions in gas phase, solid and in solution using Car Parrinello\cite{agarwal2012ab}, Born-Oppenheimer Molecular dynamics \cite{Zheng2015}, QM/MM metadynamics\cite{petersen2009mechanism}. Some applications, for example, are allyl cyanide to pirrole isomerization \cite{Pietrucci2011}, formation of silver-chloro complexes \cite{Liu2012} and water splitting and H$_2$ evolution by Ru(II)-Pincer complexes \cite{Ma2012}. To overcome the limitation of short trajectories characteristic of quantum methods, minimum activation barriers have been reported stopping the metadynamics trajectory once the first transition is achieved and averaging over a few resulting barriers\cite{Ma2012, Sgrignani2014} or continuing a single trajectory and stopping the dynamics after one recrossing has been achieved\cite{biarnes2007conformational, ghoussoub2016metadynamics}. The lack of a good estimation of the resulting errors (due to such short trajectories) is hindering a more extended use of this important free-energy estimation method. 

Noble metal nanoclusters have attracted much attention due to their molecular-like properties and high luminescence with potential applications in catalysis, biosensing and bioimaging\cite{Omoda2021}. Silver nanoclusters both bare and ligand-stabilized have a particular ability to form diverse structural motifs and a rich variety of isomers \cite{Xie2020}. Experimental and simulated absorption spectrum of ultrasmall bare silver clusters indicates the coexistence of several isomers even at low-temperature starting at $N_a$ =6 and a transition from planar to three-dimensional for its lower energy isomer at $N_a$ =7 atoms \cite{Harb2008, Duanmu2015, Chen2013}. This transition to non-planar structures is then much faster than its gold equivalent, which is placed at $N_a$ =11 up to T=100K \cite{Walker05, Goldsmith2019}. 

How would isomerization of silver clusters depend on temperature? Could the 2D-3D transition depend on temperature and other experimental conditions?  Such questions require an estimation of the free-energy landscape of small silver clusters, which is computationally challenging to compute with traditional unbiased first-principles descriptors. Some studies on gold clusters have started to address these questions with the combination of quantum methods and enhanced sampling methods. Metadynamics applied to Au$_{12}$ clusters \cite{Santarossa2010} shows that at room temperature, there is an equiprobable mixture of isomers. Recent work \cite{Goldsmith2019} on gold clusters predicts that at T=300K with $N_a$ =8 atoms there are non-planar isomers with non-negligible probabilities competing with planar isomers.    

Our goal is to determine accurately the temperature dependent free-energy landscape of small neutral clusters with the use of \textit{ab-initio} metadynamics in a general purpose interface. These results are useful to address questions of isomerization and the influence of stabilizers like solvents or organic matter. Accurate free-energy landscapes can also be used as benchmark for classical force-field developments \cite{Evangelisti2020}.

This paper is organized as follows. In the first section, we present a new interface and its validation with a classical Lennard-Jones cluster free-energy estimation. In the second section, we show an application of the \textit{ab-initio} metadynamics method for the determination of the Ag$_5$ and Ag$_6$ free-energy difference between its lowest planar and non-planar isomer, and the corresponding minimum activation barrier. We also include in this section a discussion on \textit{ab-initio} metadynamics parameters determination. We finalize with the perspectives and conclusions of the work.

\section{\label{sub:ASE-MD}ASE-PLUMED interface and validation}

Atomic Simulation Environment (ASE) is an open-source code written in python, with an object-oriented structure that allows setting, manipulating and running atomistic simulations. A central {\it{Atoms}} object obtains the energy and forces from a calculator object that in turn can be stand-alone or in interface to many of the widely used quantum and classical atomistic simulation codes. Therefore, with the aim to perform biased simulations, we created an interface to the open-source PLUMED library \cite{bonomi2009plumed,tribello2014plumed,bonomi2019promoting} by developing a calculator called Plumed that adds the energies and forces of bias to the forces and energies of other ASE calculators, which can be selected by the user. This interface between ASE and PLUMED opens the possibility to implement enhanced-sampling methods with first-principles accuracy. We note that with PLUMED, it is possible to carry out several enhanced sampling methods and post-processing analysis of trajectories.

We focus here on the implementation of well-tempered metadynamics (WT-MTD) \cite{Barducci2008}. Metadynamics is a method that adds an artificial history-dependent bias potential over a small set of collective variables (CVs) \cite{laio2002escaping}, {\bf{s}}. Typically, the bias is accumulated as the sum of Gaussians centered along the CV trajectory. This pushes the system to explore different configurations and enhanced the sampling. In particular, in WT-MTD the bias potential at time $t$ is
\begin{equation}\label{equ:bias}
    V_B ({\bf{s}}, t) = \sum_{t'=\tau, 2\tau,...}^{t'<t}W e^{-\frac{\beta \hspace{0.1cm} V_B({\bf s}, \hspace{0.1cm}t')}{\gamma}}\hspace{0.1cm} e^{-\sum_i\frac{[s_i\hspace{0.1cm} - \hspace{0.1cm}s_i(t')]^2}{2\sigma_i}}~,
\end{equation}
where $W$ is the initial height of the bias, $\tau$ is the time between deposited Gaussians, $\beta$ is the inverse of k$_B$T, $\gamma$ is a bias factor and $\sigma_i$ is the width of the Gaussians for the $i$-th CV in {\bf s}. Note that the first exponential decreases the height of the deposited Gaussians where previous bias energy has been added. This reduction of the Gaussian height reduces the error and avoids exploration towards high free energy states that are thermodynamically irrelevant. The rate at which the magnitude of the new added bias decreases is regulated by the bias factor $\gamma$: the lower the bias factor, the faster the decrease. The last exponential is a product of Gaussians in the direction of CV $i$ with width $\sigma_i$ centered at the CV value at time $t'$. In this way, the system's dynamics is enhanced, enabling it to explore different conformations. 

For a sufficiently long exploration of the conformational space, it is possible to extract the free-energy landscape  over the CVs ($F({\bf s})$) using the information from the bias potential\cite{Barducci2008}
\begin{equation}
    \lim_{t\rightarrow \infty} V_B ({\bf{s}}, t) = -\frac{(\gamma -1)}{\gamma} F({\bf s})~.
\end{equation}
This is the main advantage of MTD-derived methods.

As an accuracy test for our new Plumed calculator, we used a tutorial from the PLUMED documentation\cite{tribello2014plumed} as a benchmark system. This consists of WT-MTD/Langevin simulations for a simple system formed by seven atoms with Lennard-Jones (LJ) interactions in a planar space. The LJ cluster has several stable isomers, which can be distinguished with the CVs of second and third central moments of the distribution of the coordination numbers (labeled by SCM and TCM respectively). The nth central moment $\mu_n$ of the $N_a$-atoms cluster is defined as

\begin{equation}
   {\mu_n} = \frac{1}{N_a} \sum_{i=1}^{N_a} \left( {X}_{i} - \left< {X} \right> \right)^n ~,
\end{equation}
where $X_i$ is the coordination number of the $i$-th atom:
\begin{equation}
    X_i= \sum_{j\ne i}\frac{1-(r_{ij}/d)^8}{1-(r_{ij}/d)^{16}}~, 
    \label{eq:cnatomi}
\end{equation}
with $r_{ij}$ the distance between atoms $i$ and $j$, and $d$ a reference parameter. We used LJ dimensionless reduced units. The parameters of the simulation are $d=1.5$, $k_\text{B}T=0.1$, friction coefficient fixed equal to 1, initial bias height of 0.05, Gaussian's width of $0.1$ (for both CVs), and a bias factor of $5$.

\begin{figure}[h]
\includegraphics[width=8cm]{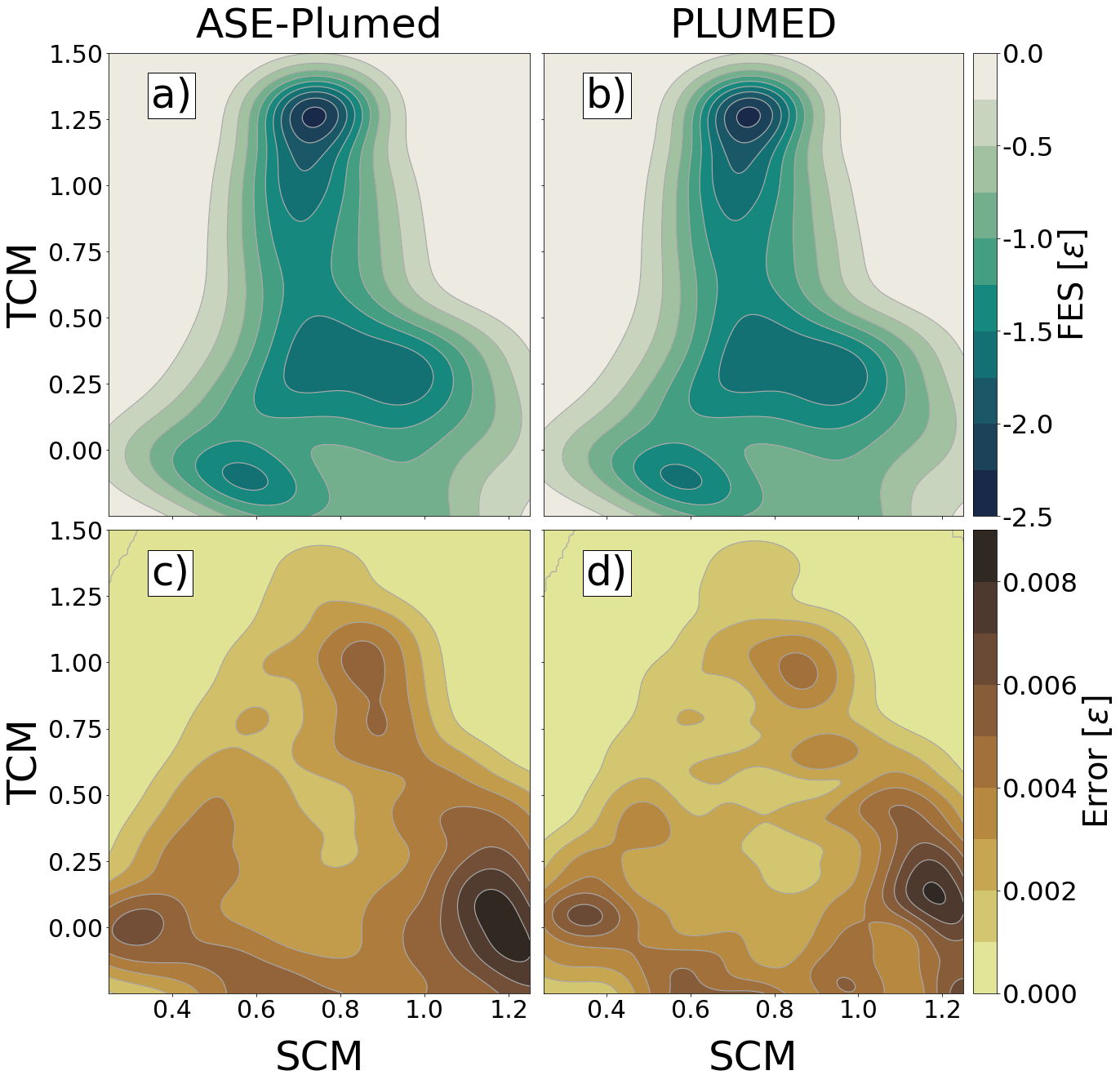}
\caption{Comparison of the free-energy landscape of seven atoms with Lennard-Jones interactions for PLUMED standalone and the new ASE-Plumed calculator using WT-MTD. (a) and (b) are the averaged free-energy landscapes in energy LJ units, over 121 trajectories, as function of CVs second and third central moments of the distribution of the coordination number. The standard error of the free-energy landscapes is shown in (c) and (d) for ASE-Plumed interface and PLUMED, respectively.}
\label{fig:test_interface}
\end{figure}

For this system, we compared the free-energy obtained by PLUMED as a stand-alone code and the free-energy estimated when using our new Plumed calculator that adds a bias force to a LJ-force calculator in ASE. For both cases, we ran 121 independent trajectories of WT-MTD, starting from the same configuration and random initial velocities, of duration 10$^6$ steps. In Fig.  \ref{fig:test_interface} a and b, we show the average free-energy surface as a function of the two CVs, for the new ASE-Plumed calculator and PLUMED alone, respectively. The free-energy error is the standard error of the 121 replicas in each grid point (\textit{i.e.}, the standard deviation over the square root of the number of simulations). These are shown in \ref{fig:test_interface} c and d. The results show that the ASE-Plumed calculator performs well, since its average free-energy landscape converges to the same values (within error) as the results from standalone PLUMED. The differences between PLUMED and ASE (for example, different random number generators) can be seen as a different error pattern in the figure. However, the error range is equal between both codes, as expected.

After testing the interface, we move into the problem of studying the temperature-dependence of the free energy landscape of small silver clusters.

\section{Free-energy landscape of small silver clusters}

In the following sections, we present the DFT and WT-MTD setup for Ag$_5$ and Ag$_6$ cluster simulations, the free-energy surfaces and results extracted at different temperatures. 

\subsection{DFT parameters}
Small neutral silver clusters have planar low-energy isomers, and as the number of atoms increases, the 3D isomers get closer in energy to the lowest energy isomer until Ag$_7$, where the trend changes and the lowest energy configuration is a 3D structure \cite{Duanmu2015, Chen2013}. 

\begin{figure}
\includegraphics[width=8cm]{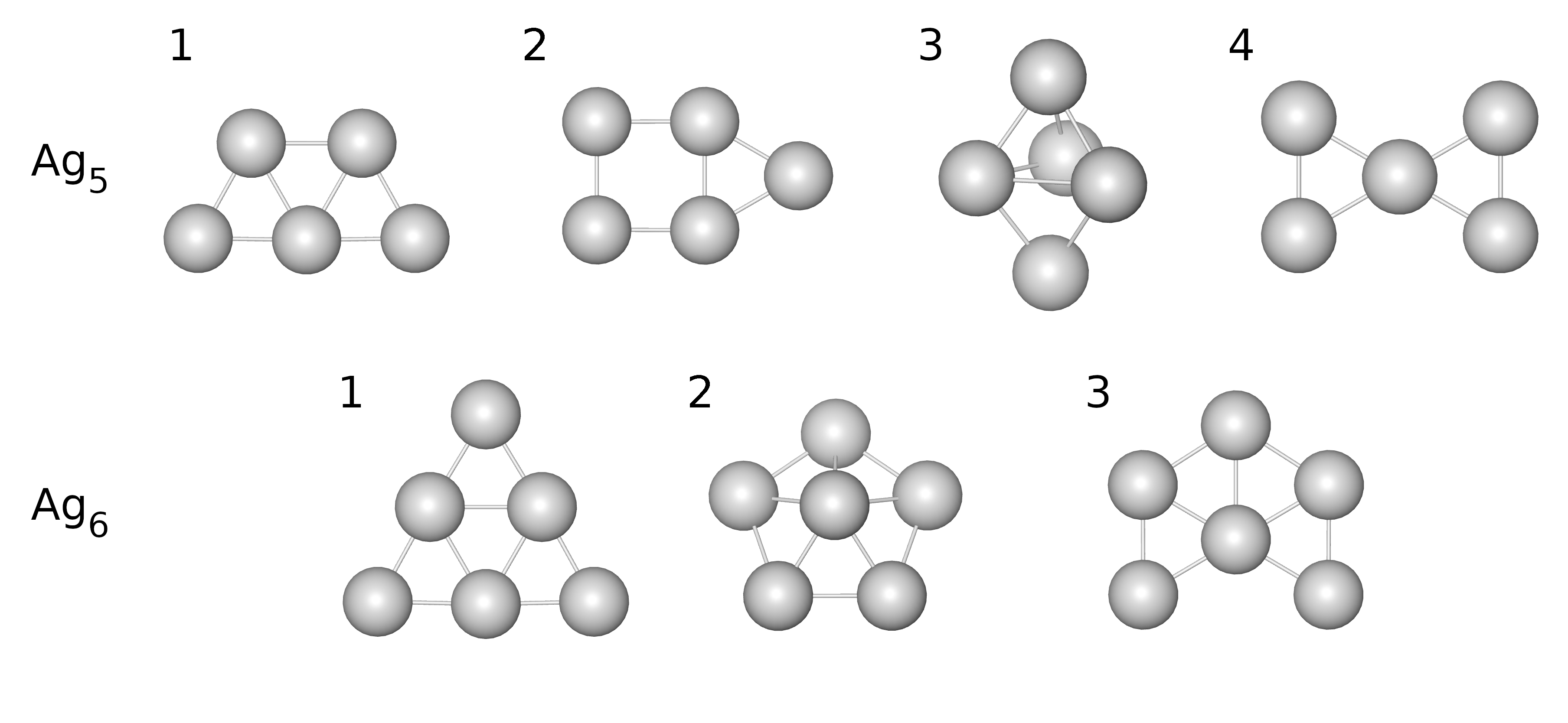}
\caption{\label{fig:isomers} Isomers of Ag$_5$ cluster in first row. 1) trapezoid, 2) edge-capped, 3) bipyramidal and 4) bow-tie. Isomers of Ag6 cluster in the second row. 1) triangular, 2) pyramidal and 3) incomplete hexagon.}
\end{figure}

In the case of Ag$_5$ cluster, experimental studies with Raman and optical photoabsorption spectroscopy agree that the lowest energy isomer of this system has a planar trapezoidal shape  \cite{Haslett1998}. Isomers of Ag$_5$ have been studied by means of computational methods such as Hartree-Fock, coupled-cluster CCSD(T) and Density Functional Theory (DFT)\cite{BonacicKoutecky1993, Duanmu2015, Chen2013}. All computational methods predict a 3D bipyramidal isomer which is about 0.4-0.5 eV higher in energy than the lowest energy state. Various methods predict other planar isomers with energies that are more strongly method-dependent. There is a planar isomer (edge-capped square) predicted by PBE, N12 and TPSS exchange-corrrelation functionals to be between the trapezoid isomer and the three-dimensional bipyramidal isomer, but not found with CCSD(T) method. Moreover, a planar isomer denoted bow-tie is found by CCSD(T) method\cite{Chen2013} at the same energy of the three-dimensional isomer but other DFT functionals, like PBE and N12, place bow-tie isomer energy 0.2-0.3 eV below the three-dimensional bipyramidal isomer energy\cite{Duanmu2015}. In Fig.  \ref{fig:isomers} (top), the trapezoidal lowest isomer is isomer 1 and the 3D bipyramidal is represented as isomer 3. The edge-capped square isomer is isomer 2 in Fig.  \ref{fig:isomers} (top).  Bow-tie isomer is isomer 4 in Fig.  \ref{fig:isomers} (top). 

Likewise, Ag$_6$ cluster has been studied experimentally and computationally, using DFT\cite{Duanmu2015} and CCSD(T) method\cite{Chen2013}. All computational methods suggest a triangular-planar isomer as configuration of minimum energy, followed by a 3D pyramidal isomer with a difference of energy  of 0.1-0.2 eV. A third isomer (planar incomplete hexagon) is predicted with an energy of 0.3 eV with respect to the minimum energy configuration. Experiments of absorption spectrum suggests a possible mixture of triangular and pyramidal isomers, although the difference of energies makes the presence of the pyramidal isomer not really favorable according to the zero energy analysis \cite{Lecoultre2011, Harb2008}. Fig.  \ref{fig:isomers} (bottom) shows the isomers of Ag$_6$: the lowest in energy, triangular isomer, labeled as isomer 1; the next stable configuration, 3D pyramidal, labeled as isomer 2; and the third isomer, incomplete hexagon.

Table \ref{tab:table1} summarizes the potential energies of all isomers discussed in this section. We include reference values (columns CCSD(T) and N12), energies computed here with finite difference basis (columns PBE and TPSS) and energies computed with the PBE exchange-correlation but with the faster and less accurate LCAO pvalence basis (column PBE-LCAO-PVAL).

Comparing Ag$_5$ PBE and CCSD(T) columns, we observe that PBE gives an overestimation of the 2D-3D energy difference by about 0.1 eV which is reduced with the use of the basis pvalence. On the opposite direction, for Ag$_6$, DFT-PBE gives a good estimation of the 2D-3D energy difference which then gets underestimated once the LCAO pvalence replaces the finite-difference method. This result would point to an effect of the PBE delocalization of electronic density, which is then slightly corrected by the use of the localized atomic basis LCAO pvalence, but such cancellation is only beneficial in the case of Ag$_5$. 

In the column TPSS of Table \ref{tab:table1}, we report optimization of isomers with TPSS exchange-correlation functional and finite difference basis. As it was reported earlier for gold clusters\cite{Ferrighi09}, this functional gives a  good accuracy and reproducing the order and energies of CCSD(T) calculations. We suggest its use in future simulations, although it was out of reach for the computational resources used in this work. Here, we used PBE exchange-correlation functional with the LCAO basis which gives the right 2D-3D ordering with very good efficiency. 

\begin{table}[h]
\caption{\label{tab:table1} Silver cluster isomer energies in eV relative to the lowest isomer with different methods. We include energies obtained in with coupled cluster method, DFT functionals like N12, PBE and TPSS. With the functional PBE we also include the predicted energies with the faster basis LCAO p-valence used in this work.}
\begin{ruledtabular}
\begin{tabular}{cccccccccc}
System & Isomer & Symmetry & Dimension &CCSD(T)\cite{Chen2013}& N12 \cite{Duanmu2015}& PBE & PBE-LCAO-PVAL&TPSS\\
\hline
Ag$_5$& 1 & C$_{2v}$& 2 & 0 & 0 & 0 & 0 & 0 \\
&2 & C$_{2v}$& 2 & - & 0.27 & 0.22  & 0.26 &0.27 \\
&3 & C$_{2v}$& 3 & 0.43 & 0.53 & 0.55  & 0.40 &0.46\\
&4 & D$_{2h}$& 2 & 0.46 & 0.36 & 0.39  & 0.53 & 0.43\\
Ag$_6$ & 1 & D$_{3h}$& 2 & 0 & 0 & 0 & 0 & 0\\
&2 & C$_{5v}$& 3 & 0.20 & 0.25 & 0.23 & 0.09 & 0.21\\
&3 & C$_{2v}$& 2 & 0.30 & 0.29 & 0.28 & 0.27& 0.28
\end{tabular}
\end{ruledtabular}
\end{table}

In order to take in account variations due to entropic effects and temperature, we apply WT-MTD to these two systems using the CVs described below.

\subsection{Collective variables for WT-MTD}

CVs are functions of the coordinates of the atoms that help to extract biophysical properties, separate relevant metastable states and reduce the dimensionality. CVs project the multi-dimensional system onto a small set of relevant and, in most cases, interpretative degrees of freedom. CVs are used in MTD and other enhanced-sampling methods (such as umbrella sampling \cite{torrie1977nonphysical}) to add bias to the system. To guarantee a convergence, the CVs have to be chosen carefully, such that they differentiate the isomers and the transition states. 

We studied several choices of CVs for the Ag$_5$ and Ag$_6$ systems. In particular, we focused on the coordination number $C$ and radius of gyration $R$, which were previously used for studying the conformations of Au$_{12}$ cluster \cite{Santarossa2010}. The average coordination number is 
\begin{equation}
    C= \sum_{i=1}^{N_a}X_i,
\end{equation}
where $X_i$ is defined in Eq. \ref{eq:cnatomi} and the reference distance $d$ was set to 2.8$\text{\AA}$ to include all first neighbors distances in the silver isomers. This CV measures the number of bonds in the system. 

The radius of gyration is
\begin{equation}
    R= \left(\frac{\sum_i^N |r_i - r_{CM}|^2}{N_a}\right)^{1/2}~,
\end{equation}
where $r_i$ is the position of atom $i$, $r_{CM}$ is the center of mass of the cluster and $N_a$ is the number of atoms of the cluster. This CV gives information about how disperse the system is with respect to the center of mass. $C$ and $R$ enable extracting information about the shape of the cluster and permit differentiating the free-energy minima found by DFT optimization, which are expected to be metastable states in the free-energy landscape. 

To evaluate more precisely the suitability of this set of CVs, we performed unbiased of MD in ASE for 10000 steps with a 5 fs time-step using the Born-Oppenheimer Approximation. The electronic distribution was obtained with LCAO-pvalence basis in a cell of 16 \AA, using the GPAW calculator \cite{mortensen2005real}. The temperature was controlled with a Berendsen thermostat at 10 K with a $\tau_t$ of 50 fs (that will be the same setup used in WT-MTD for a wider range of temperatures). Starting from the trapezoid state and bipyramidal state, we observe that the form of the basins in the space of these collective variables were tilted ellipsoids in the unbiased MD (Fig. S1 and Fig. S2 in Supplementary). Therefore, there are regions of the space that are thermodynamically irrelevant. To avoid enhancing the exploration toward these regions, we created a new set of CVs (CV1 and CV2) that are a rotation of $C$ and $R$, over which we could easily apply a wall. The rotated CVs are defined as
\begin{eqnarray}
CV1 = 0.99715 \hspace{0.2cm}C - 0.07534\text{\AA}^{-1} \hspace{0.2cm}R \\
CV2 = 0.07534 \hspace{0.2cm}C + 0.99715\text{\AA}^{-1} \hspace{0.2cm}R .
\end{eqnarray}

Using this CV setup for WT-MTD, we added walls using repulsive semi-harmonic potentials that act when CV1 is lower than 5 with harmonic constant 10 eV and when CV2 is greater than 3 with harmonic constant 50 eV for Ag$_5$ (dashed lines in Fig. \ref{fig:feses}). In the case of Ag$_6$, the walls repel values of CV1 lower than 8 with harmonic constant 10 eV and values of CV2 greater than 3.3 with harmonic constant 50 eV. We note that all isomers of Ag$_5$ and Ag$_6$ appear discriminated in the space of the CV1 and CV2. Moreover, this combination of the coordination number and radius of gyration in the CV2 variable is a good collective variable that allows to represent the FE of Ag$_6$ along a 1D profile (as will be described below).

\subsection{WT-MTD parameters}

\begin{figure}[h]
\includegraphics[width=8cm]{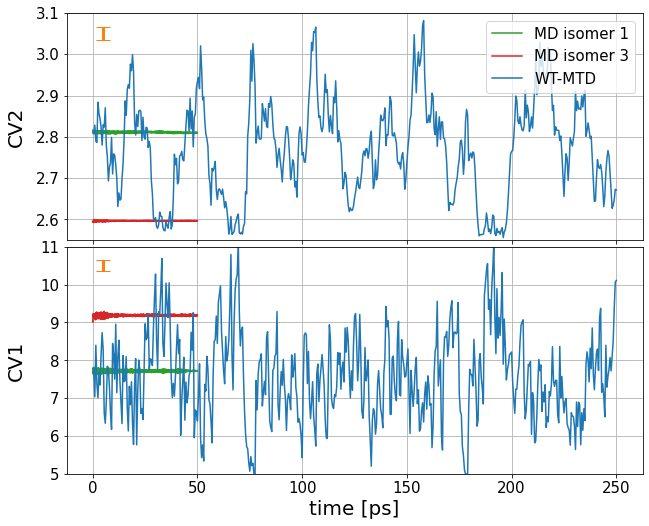}
\caption{Example of the evolution of CV1 and CV2 in MD without bias and with biased WT-MTD at $T=10K$. Red and green lines are unbiased MD simulations starting from Ag$_5$ states 1 and 3 from Fig. \ref{fig:isomers}. The maximum variation range was used to set the Gaussian width $\sigma_{CV1}$ and $\sigma_{CV2}$ (shown as a bars). Blue lines represent the evolution of the collective variables in the WT-MTD.}
\label{fig:MD_noBias-WT-MTD}
\end{figure}

We used the unbiased MD trajectories to determine the optimal parameters for the WT-MTD simulation. By monitoring the CVs as a function of time, we can estimate the MTD Gaussian width, which should approximate the amplitude of the CV at each minimum (bars in Fig.  \ref{fig:MD_noBias-WT-MTD}). In other words, the Gaussian widths are on the same order as the variation of the CVs in the unbiased simulation. Therefore, we choose the values of $\sigma_{CV1}$ and $\sigma_{CV2}$ fixed to 0.3 and 0.03, respectively.

For choosing the other WT-MTD parameters, we performed several simulations using classic metadynamics to obtain an idea of the barrier height between different states, and extracting an optimal setup. From this exploration step, we decided to fix the initial height to 0.3 eV for Ag$_5$ and 0.2 eV for Ag$_6$. The bias factor was fixed in 500, 100 and 50 for the temperatures 10, 100 and 300 K, respectively, in such a way that enabled the system to jump from the deepest minimum, but the Gaussians decrease sufficiently fast to achieve convergence in the simulated steps.

\subsection{Ag$_5$ FES from low to room temperature}

Using the paramters and CVs described above, we performed WT-MTD on Ag$_5$ clusters for 11 independent replicas and 50000 steps, resulting in a total of 250 ps. We obtained a clear difference in the exploration of the configurations compared with the unbiased simulation. For WT-MTD, the CVs filled the metastable state smoothly and then many transitions between states were observed. As a consequence, the system explored a large range of values in comparison with the maximum range covered by the simulations without bias (Fig.  \ref{fig:MD_noBias-WT-MTD}). This demonstrates that the free-energy landscape is being filled by the bias potential and that the system is recrossing the relevant metastable states. Importantly, we remark that it is not enough to find one single transition because the free-energy reconstruction will be poor. Therefore, the simulations ran until the error (calculated using $N$ trajectories) was in the order of tens of meV, that condition usually implies more than 4 transitions between minima. 

\begin{figure}[h]
\includegraphics[width=16cm]{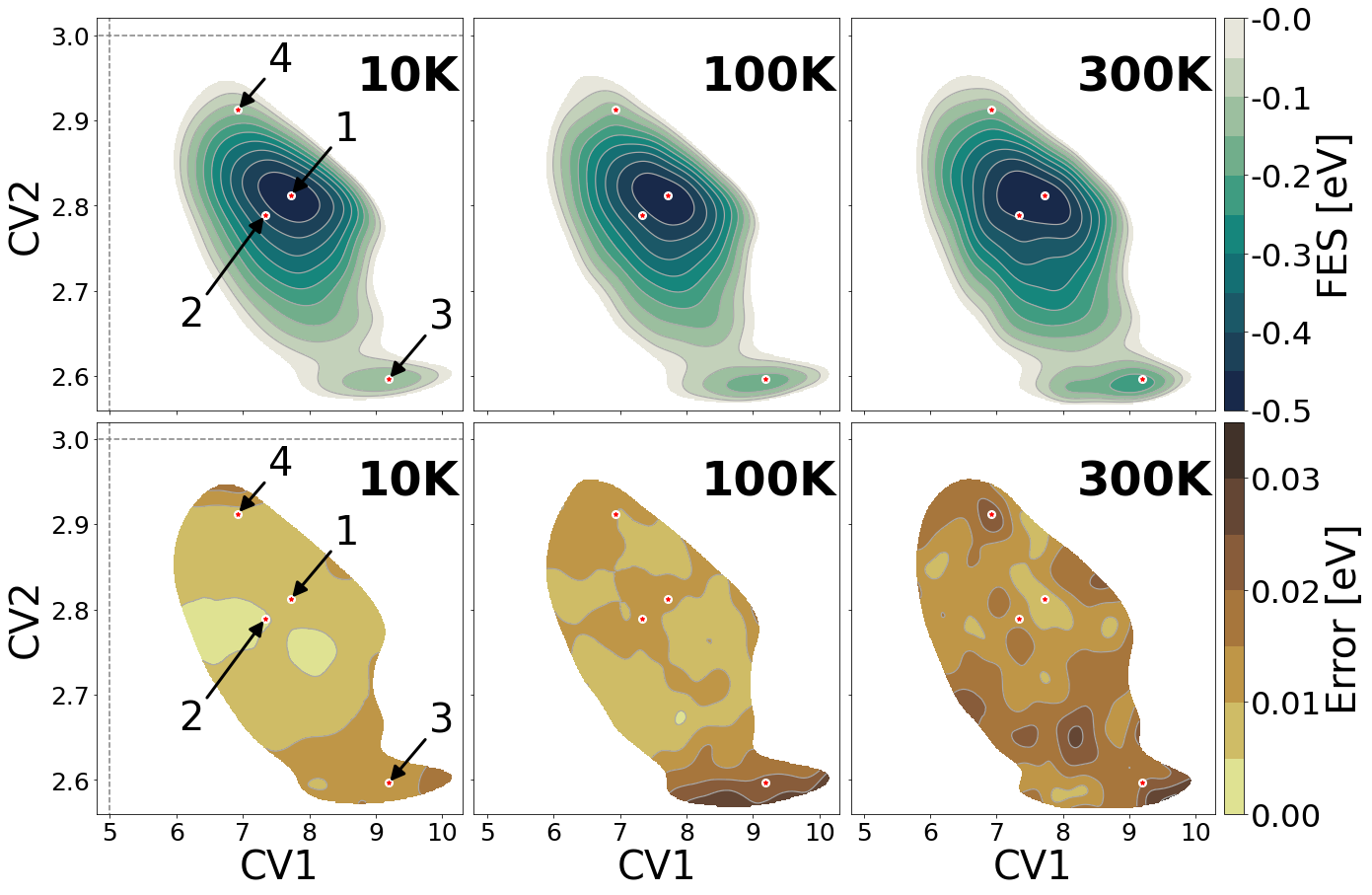}
\caption{Free-energy surface of Ag$_5$ and error obtained in the space of the CVs CV1 and CV2 at temperatures 10K, 100K and 300 K. Dashed lines in 10 K are the limit of the lower (vertical) and upper (horizontal) walls that avoid an exploration towards high energy regions. The level curves are placed each 0.05 eV for the FES and 0.005 eV for the error. The positions of isomers (Fig. \ref{fig:isomers}) are shown as dots. }
\label{fig:feses}
\end{figure}

We estimated the free-energy surface in the space of CV1 and CV2 as the average of the FE for the 11 trajectories for three different temperatures: 10K, 100K and 300K (Fig.  \ref{fig:feses} (top)). For all temperatures, the free-energy landscape contain just two minima, although four minima are obtained from optimization. At these temperatures, states 2 and 4 (shown in Fig. \ref{fig:isomers}), corresponding with the edge-capped square and bow-tie isomers, are just saddle points that belong to state 1 (the trapezoid isomer). Therefore, only states 1 and 3 (shown in Fig. \ref{fig:isomers}) are representative configurations of stable isomers. With increasing temperature in Fig. \ref{fig:feses}, the general form of the free energy is conserved, but both minima are more populated when the temperature is larger, as expected since the system has more thermal energy that enables it escaping from the local minimum and occupying other states. In Fig. \ref{fig:feses} (bottom), we present the standard error calculated as the standard deviation in each grid point over the root square of the number of replicas, $N$. It is on the order of the tens of meV at most, but remains lower around the lowest free-energy regions, namely, in the regions close to the minima, suggesting a good reliability of the FE reconstructions.

\begin{figure}
\includegraphics[width=8cm]{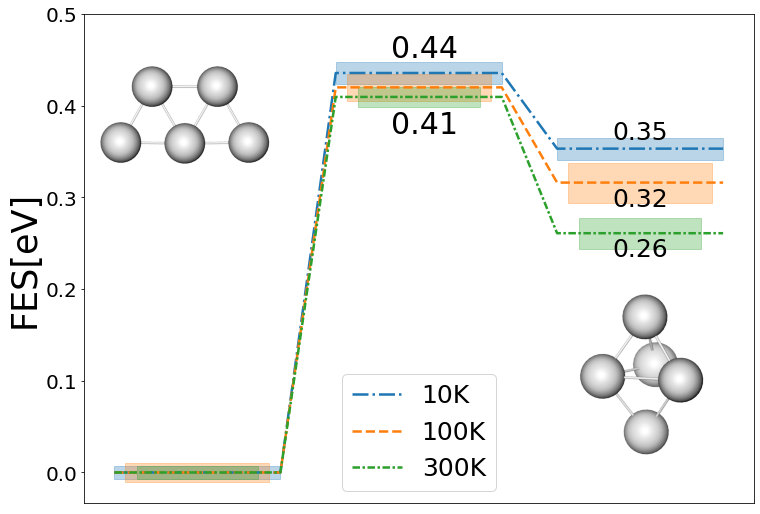}
\caption{Minimum activation barrier and free-energy of the state 3 (shown at right) relative to the free-energy of state 1 (shown at left) at temperatures 10, 100 and 300 K for Ag$_5$ cluster.}%
\label{fig:relatives}
\end{figure}

For low temperatures, we expect only a small variation between the computed zero temperature energy obtained via DFT optimization and the 10K free energy obtained with \textit{ ab initio} metadynamics because of the underlying assumptions of fixed Boltzmann statistics in metadynamics and Born-Oppenheimer electron-nucleus decoupling in DFT. We use therefore the 10K FES values as convergence check inside our trend study, which converge to the expected values. We note that quantum effects that are not valid under these assumptions will not be captured in our simulations. The effect of increasing the temperature is given by a decrease in the minimum activation barrier and the free-energy difference between minima between the planar and non-planar isomers. These results are shown in Fig.  \ref{fig:relatives}, finding that the transition state decreases approximately 0.03 eV and the free energy difference decreases by 0.09 eV from 10K to 300K. 

A more dramatic change is captured when calculating the relative population of the basins using the Boltzmann factor. We define a basin as the region where the free-energy surface is less than the value of the minimum activation barrier (Fig.  \ref{fig:relatives}). Then, we obtained the probability of each state by integrating the Boltzmann factor over the correspondent basin, \textit{i.e.} the probability of state $1$ is $P_1 = \int_{1} exp(-\beta F(\textbf{s})) d\textbf{s}$ where $\beta=1/(k_B T)$ and $F(\textbf{s})$ is the free-energy at $\textbf{s}$. Interestingly, the probability associated to all the non-planar isomers is negligible for all temperature ranges (even $300K$), namely, the probability to find a planar configuration is 100\% for Ag$_5$.

\begin{figure}[h]
\includegraphics[width=12cm]{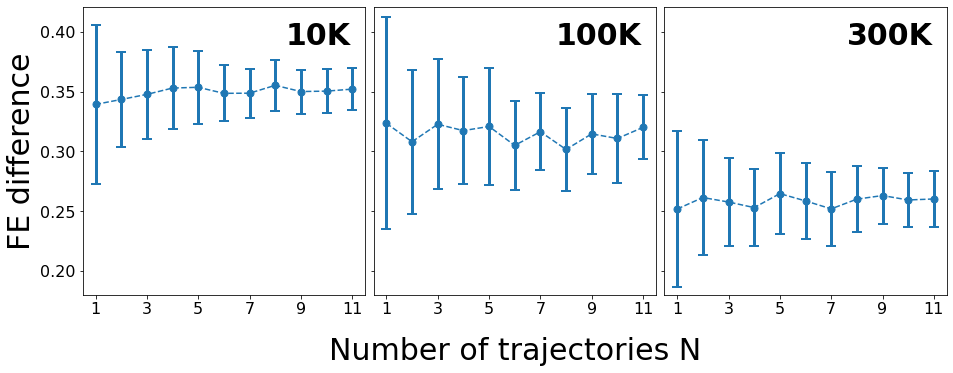}
\caption{Bootstrap analysis of the difference between the Ag$_5$ free-energy minima, state 1 and state 3, in terms of the number of bootstrap samples. The mean (dots) and standard error (bars) are shown. As the number of samples increases, the error decreases.}
\label{fig:differences_bootstrap}
\end{figure}

 Using statistical bootstrapping with 50 resamples, we explored how many independent simulations are required to extract an error by varying the number of samples in each resampling. This gives a notion of how  the predicted result changes as a function of the number of simulated replicas. In Fig.  \ref{fig:differences_bootstrap}, we show the mean value (dots) and the standard deviation (bars) of the difference in free-energy between the isomers 1 and 3 of Ag$_5$. This result demonstrates the importance of running at least 4 replicas for obtaining a reliable free-energy difference estimate. We note that when using only one metadynamics simulation, the results can significantly change, even up to 0.15 eV, which is a large variation compared with the value of this observable. This demonstrates the importance of considering several replicas for relatively short simulations. We note that the exact convergence rate will depend also on the complexity of the particular system, the simulation length and WT-MTD setup.  

\subsection{Ag$_6$ FES from low to room temperature}

\begin{figure}
\includegraphics[width=16cm]{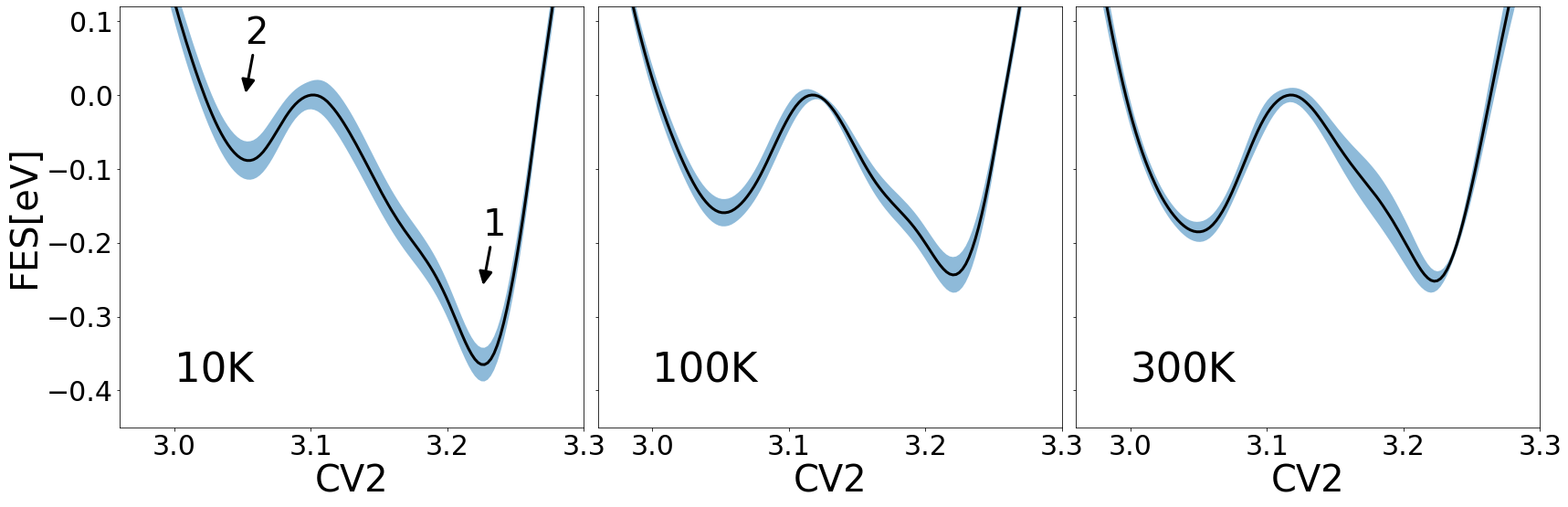}
\caption{Free-Energy profile of Ag$_6$ at 10, 100 and 300 K along the collective variable CV2 with CV1 integrated out. The shaded region shows the standard error. Arrows point the values of CV2 for the isomers 1 and 2 in Fig. \ref{fig:isomers}.}%
\label{fig:fesAg6}
\end{figure}

We also studied the free-energy landscape of the Ag$_6$ cluster, running 4 independent trajectories with 136000 steps, resulting in a total of 680 ps. The cluster has three stable isomers according to the optimization analysis (Table \ref{tab:table1} and Fig. \ref{fig:isomers}). In Supplementary Fig. S3, we show the FES along both CVs at the simulated temperatures. We note that only the isomers 1 and 2 are stable states of the free-energy landscape. The incomplete hexagon isomer, the third Ag$_6$ isomer, appears as part of the basin of the isomer 1. From that figure, it is clear that states 1 and 2 are properly separated along CV2. Therefore, integrating CV1 enables a clear representation of a free-energy profile along CV2 (\textit{i.e.} $\exp(-\beta F(CV2)) = \int  \exp(-\beta F(CV1,CV2))\, dCV1.$). The averaged profile over the 4 trajectories, and a shaded region representing the standard error are shown in Fig. \ref{fig:fesAg6}. For the three cases, the standard error remains lower than 0.04 eV.

From 10 K to room temperature, the free-energy difference decreases approximately by 0.2 eV, but the barrier with respect to the global minima decreases by only 0.02 eV. In terms of probability, however, the change is drastic. At room temperature the probability of non-planar isomer reaches 10\% which shows that the system reaches a new equilibrium where planar and non-planar isomers are competing. This is a marked difference between Ag$_5$ and Ag$_6$, and it is in accordance to an observed change in optical spectrum experiments \cite{Lecoultre2011}.

\newpage
\newpage
\section{Conclusion}
We developed a new calculator for ASE called Plumed, which can be used for running simulations of enhanced-sampling methods by patching the open-source code ASE and the plugin PLUMED. This calculator was tested with a simple system of seven LJ atoms as benchmark. This ASE-PLUMED interface was used for studying Ag$_5$ and Ag$_6$ clusters at different temperatures. We found crucial thermal effects over the Ag$_6$ system, which changes from a planar-dominated population at low temperatures to a state with mixture of planar and non planar isomers at room temperature. Because no changes in population are found in Ag$_5$ at the same temperature range, it follows then that Ag$_6$ is the smallest silver cluster with a 2D-3D isomer equilibrium at room temperature.

As an important result, we studied the convergence of the WT-MTD technique, showing that when considering trajectories with few recrossings using just one replica could generate unreliable results but averaging over several independent replicas decreases significantly the error. All-in-all, we foresee that the ASE-PLUMED interface as a general purpose tool for enhanced sampling simulations  having the accuracy of first principles electronic structure methods will expand  this method use in the material's science community.

\section{Data Availability Statements}
The data that supports the findings of this study are available within the article and its supplementary material. The code developed in this study is openly available in gitlab (https://gitlab.com/ase/ase). A tutorial is also available at ASE webpage (https://wiki.fysik.dtu.dk/ase).

\begin{acknowledgments}
P.C. has been supported by MinCiencias, University of Antioquia (Colombia), and the Simons Foundation (USA). D. S, C. P and O.L-A have been supported by Minciencias and University of Antioquia (Colombia).
\end{acknowledgments}

\bibliography{bibliography}

\end{document}



\title{\centering Supplementary Information\\
Ab initio metadynamics determination of temperature-dependent free-energy landscape in ultrasmall silver clusters} 


\author{Daniel Sucerquia}
\affiliation{Biophysics of Tropical Diseases, Max Planck Tandem Group, University of Antioquia UdeA, 050010 Medellin, Colombia
}%
\affiliation{Grupo de Física Atómica y Molecular, Instituto de Física, Facultad de Ciencias Exactas y Naturales, Universidad de Antioquia UdeA; Calle 70 No. 52-21, Medellín, Colombia}

\author{Cristian Parra}%
\affiliation{Biophysics of Tropical Diseases, Max Planck Tandem Group, University of Antioquia UdeA, 050010 Medellin, Colombia
}%

\author{Pilar Cossio}
\email{pcossio@flatironinstitute.org}
\affiliation{Biophysics of Tropical Diseases, Max Planck Tandem Group, University of Antioquia UdeA, 050010 Medellin, Colombia
}%
\affiliation{Center for Computational Mathematics, Flatiron Institute, NY, USA.
}%

\author{Olga Lopez-Acevedo}
\email{olga.lopeza@udea.edu.co}
\affiliation{Biophysics of Tropical Diseases, Max Planck Tandem Group, University of Antioquia UdeA, 050010 Medellin, Colombia
}%
\affiliation{Grupo de Física Atómica y Molecular, Instituto de Física, Facultad de Ciencias Exactas y Naturales, Universidad de Antioquia UdeA; Calle 70 No. 52-21, Medellín, Colombia}

\date{\today}

\pacs{}

\maketitle 


\begin{figure}[h]
\includegraphics[width=8cm]{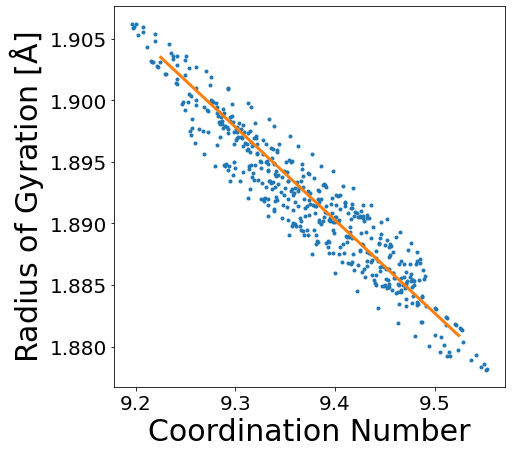}
\caption{Example of the unbiased MD trajectory at $10$ K starting from the bipyramidal state (dots) in the space of the collective variables coordination number and radius of gyration. The direction of the collective variable CV1 is showed as a solid red line and CV2 is orthogonal to it.}
\label{fig:MD_bipi}
\end{figure}

\begin{figure}[h]
\includegraphics[width=8cm]{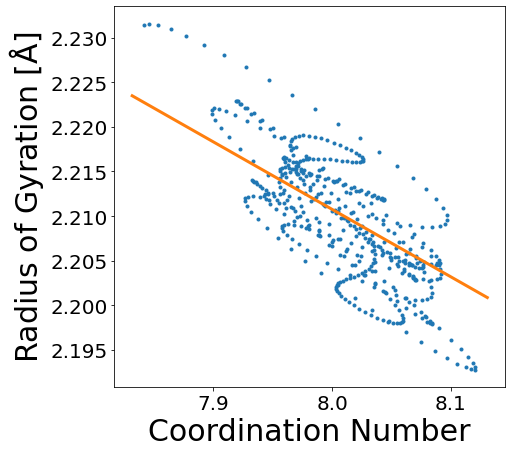}
\caption{Example of the unbiased MD trajectory at $10$ K starting from the trapezoidal state (dots) in the space of the collective variables coordination number and radius of gyration. The direction of the collective variable CV1is showed as a solid red line and CV2 is orthogonal to it.}
\label{fig:MD_bipi}
\end{figure}

\begin{figure}[h]
\includegraphics[width=15cm]{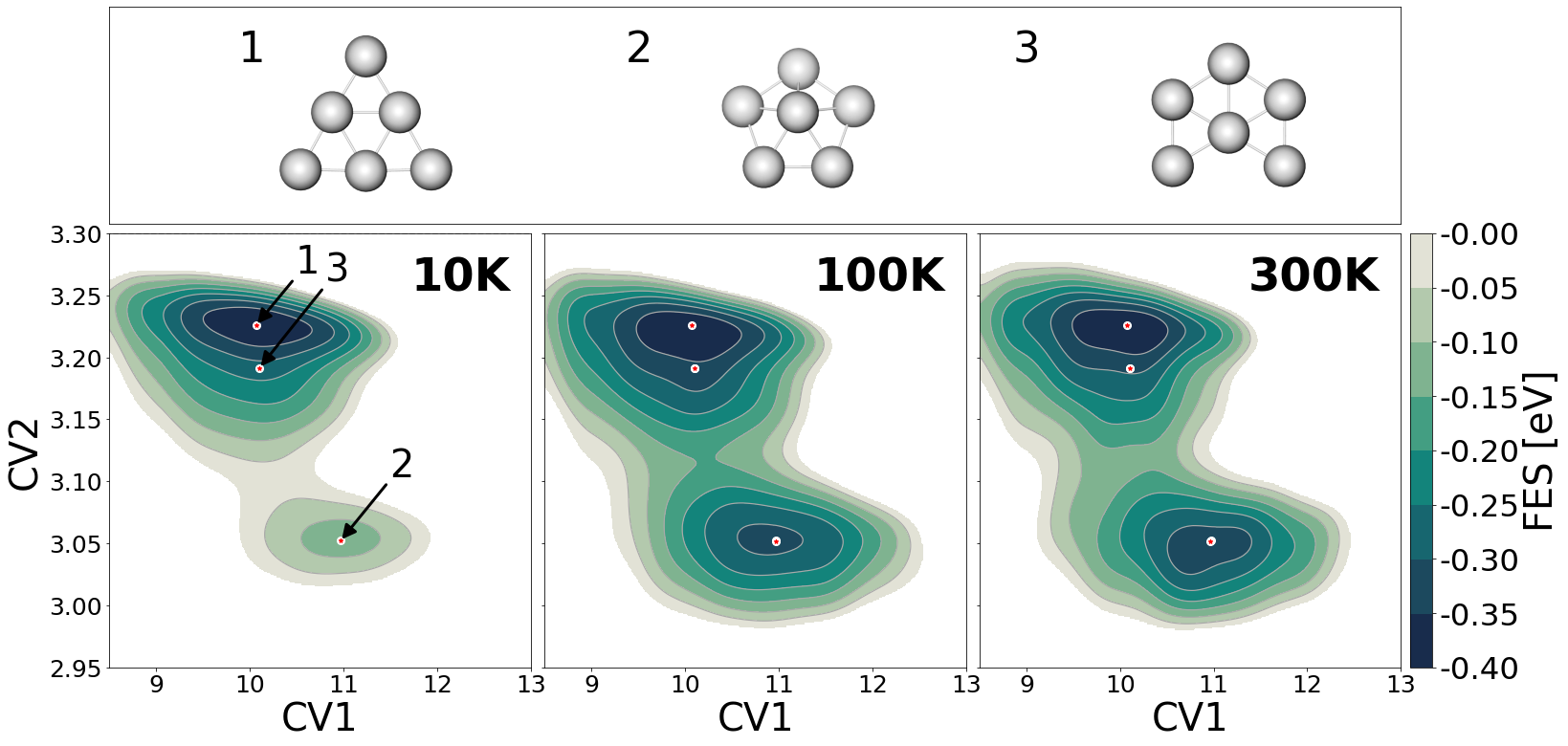}
\caption{Free-energy surface of Ag$_6$ obtained in the space of the CVs CV1 and CV2 at temperatures 10K, 100K and 300 K. The level curves are placed each 0.05 eV. The most relevant isomers (top - 1, 2 and 3) are shown as dots.}
\label{fig:Ag6_2D}
\end{figure}